\documentclass[iop]{emulateapj}
\usepackage{lscape}

\begin{document}

\shorttitle{}
\shortauthors{Falcon, Winget, Montgomery, \& Williams}

\accepted{February 2, 2010}
\submitted{December 2, 2009}

\title{A Gravitational Redshift Determination of the Mean Mass of White Dwarfs. DA Stars}

\author{Ross E. Falcon, D.E. Winget, M.H. Montgomery, and Kurtis A. Williams}
\affil{Department of Astronomy and McDonald Observatory, University of Texas, Austin, TX, 78712}
\email{cylver@astro.as.utexas.edu}

\begin{abstract}
We measure apparent velocities ($v_{\rm app}$) of the H$\alpha$ and H$\beta$ Balmer line cores for 449 non-binary thin disk normal DA white dwarfs (WDs) using optical spectra taken for the ESO \textbf{S}N Ia \textbf{P}rogenitor surve\textbf{Y} \citep[SPY;][]{napiwotzki01an}.  Assuming these WDs are nearby and co-moving, we correct our velocities to the Local Standard of Rest so that the remaining stellar motions are random.  By averaging over the sample, we are left with the mean gravitational redshift, $\langle v_{\rm g}\rangle$: we find $\langle v_{\rm g}\rangle =\langle v_{\rm app}\rangle=32.57\pm1.17$\,km s$^{-1}$.  Using the mass-radius relation from evolutionary models, this translates to a mean mass of $0.647^{+0.013}_{-0.014}$\,M$_\odot$.  We interpret this as the mean mass for all DAs.  Our results are in agreement with previous gravitational redshift studies but are significantly higher than all previous spectroscopic determinations {\it except} the recent findings of \citet{tremblay09}.  Since the gravitational redshift method is independent of surface gravity from atmosphere models, we investigate the mean mass of DAs with spectroscopic $T_{\rm eff}$ both above and below 12000\,K; fits to line profiles give a rapid increase in the mean mass with decreasing $T_{\rm eff}$.  Our results are consistent with {\it no} significant change in mean mass: $\langle M\rangle$\,$^{\rm hot}=0.640\pm0.014$\,M$_\odot$ and $\langle M\rangle$\,$^{\rm cool}=0.686^{+0.035}_{-0.039}$\,M$_\odot$.
\end{abstract}


\begin{figure*}[t!]
\plotone{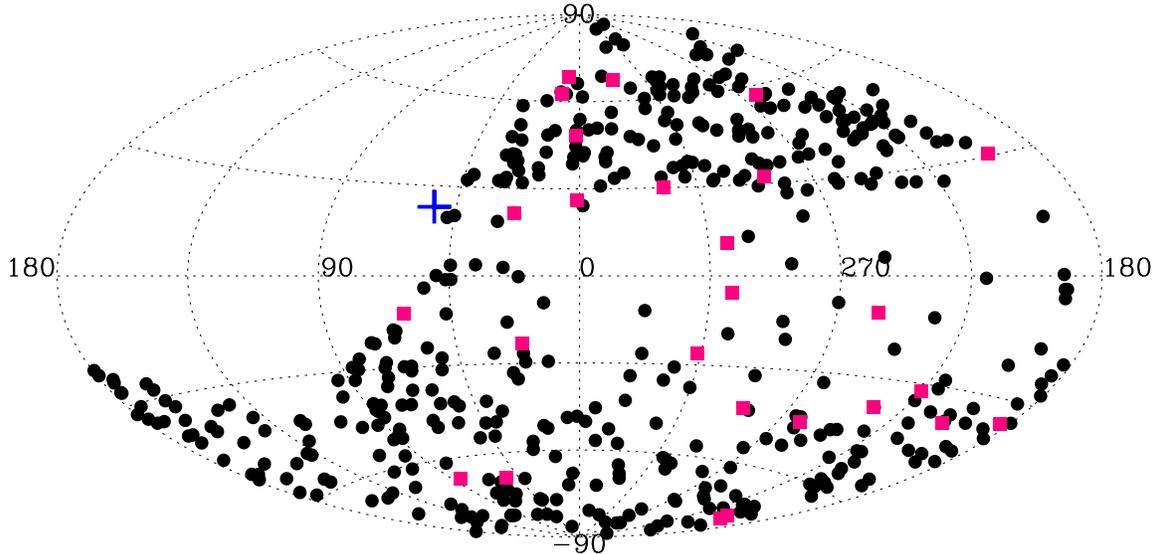}
\caption{Distribution of targets in Galactic longitude $l$ and latitude $b$.  We mark the targets in our main sample as black points and the thick disk WDs as pink squares.  We indicate the direction of the movement of the Sun with respect to the LSR \citep[{\it blue cross}\,;][]{kerr86}.  Since the observations are from the ESO VLT in the Southern Hemisphere, no targets with a declination above $+30^\circ$ are in our sample, hence the gap in the left side of the plot.\label{galplot}}
\end{figure*}

\section{Introduction}\label{intro}

Nearly all stars end their trek through stellar evolution by becoming white dwarfs (WDs).  Hence, properties of WDs can provide important information on the chemical and formation history of stars in our Galaxy, as well as on the late stages of stellar evolution.  Of these stellar properties mass is one of the most fundamental, and though there are several methods for mass determination of WDs, each has its limitations.

The most-widely used WD mass determination method involves comparing predictions from atmosphere models with observations to obtain effective temperatures ($T_{\rm eff}$) and/or surface gravities (log\,$g$).  One can then compare these quantities with predictions from evolutionary models \citep[e.g.,][]{althaus98,montgomery99}.  \citet{shipman79}, \citet{koester79}, and \citet{mcmahan89} use radii determined from trigonometric parallax measurements along with $T_{\rm eff}$ from photometry to determine masses.  Of course this technique is limited to target stars with measured parallaxes, so users of photometry have more often used observed color indices to determine both $T_{\rm eff}$ and log\,$g$ \citep[e.g.,][]{koester79,wegner79,shipman80,weidemann84,fontaine85}.  With the exception of the parallax variant \citep{kilic08}, the photometric method is seldom used in recent WD research.

Another variant of this method uses mainly spectroscopic rather than photometric observations \citep[e.g.,][]{bergeron92b,finley97,liebert05}.  With more recent large-scale surveys, such as SPY (see Section \ref{obs}) and the Sloan Digital Sky Survey \citep[SDSS;][]{york00}, the comparison of observed WD spectra with spectral energy distributions of theoretical atmosphere models has become the primary WD mass determination method, yielding masses for a large number of WDs \citep[e.g.,][]{koester01,madej04,kepler07}.

When applied to cool WDs ($T_{\rm eff}\lesssim12000$\,K), however, the reliability of this primary method breaks down: a systematic increase in the mean log\,$g$ for DAs with lower $T_{\rm eff}$ has repeatedly shown up in analyses \citep[e.g.,][]{liebert05,kepler07,degennaro08}.  This ``log\,$g$ upturn'', discussed thoroughly by \citet{bergeron07} and by \citet{koester09a}, is generally believed to reflect shortcomings of the atmosphere models $-$ specifically our understanding of the line profiles $-$ rather than a real increase in mean mass with decreasing $T_{\rm eff}$ as an increasing mean log\,$g$ would imply.

Other mass determination methods that are independent of atmosphere models include the astrometric technique \citep[e.g.,][]{gatewood78} and pulsational mode analysis \citep[e.g.,][]{winget91}.  Unfortunately, neither of these methods are widely applicable to WDs.  The former requires stellar systems with multiple stars, and the latter is limited to WDs and pre-white dwarfs which lie in narrow $T_{\rm eff}$ ranges of pulsational instability.

Another method that is mostly atmosphere model-independent uses the gravitational redshift of absorption lines; this is the one that will be the focus of this paper.  The difficulty in disentangling the stellar radial velocity shift from the gravitational redshift has caused this method to only be used for WDs in common proper motion binaries or open clusters \citep{greenstein67,koester87,wegner91,reid96,silvestri01}.  The simplicity of this method, however, prompts us to extend the investigation beyond those cases.

In this paper, we will make two main points: (1) by using a large, high-resolution spectroscopic dataset, we can circumvent the radial velocity-gravitational redshift degeneracy to measure a {\it mean} gravitational redshift of WDs in our sample and use that to arrive at a mean mass; and (2) since the gravitational redshift method has the advantage of being independent of surface gravity from atmosphere models, we can use it to reliably probe cool DAs ($T_{\rm eff}\lesssim12000$\,K), thus providing important insight into the ``log\,$g$ upturn problem'' as groups continue to improve upon those models \citep[e.g.,][]{tremblay09}.

\section{Gravitational Redshift}\label{red}

In the weak-field limit, the general relativistic effect of gravitational redshift ($z$) can be understood, classically, as the energy ($E$) lost by a photon as it escapes a gravitational potential ($\Phi$) well:
\begin{equation}
z=\frac{-\Delta E}{E}=\frac{-\Phi}{c^2}.
\end{equation}
The fractional change in energy can be rewritten as a fractional change in observed wavelength ($-\Delta E/E=\Delta\lambda/\lambda$).  In our case, the gravitational potential is at the surface of a WD of mass $M$ and radius $R$.  In terms of a velocity, the gravitational redshift is
\begin{equation}\label{v_g}
v_g=\frac{c\Delta\lambda}{\lambda}=\frac{GM}{Rc}
\end{equation}
where $G$ is the gravitational constant, and $c$ is the speed of light.

For WDs, $v_{\rm g}$ is comparable in magnitude to the stellar radial velocity $v_{\rm r}$, both of which sum to give the apparent velocity we measure from absorption lines: $v_{\rm app}=v_{\rm g}+v_{\rm r}$.  These two components cannot be explicitly separated for individual WDs without an independent $v_{\rm r}$ measurement or mass determination.

The method of this paper is to break this degeneracy not for individual targets but for the sample as a whole.  We make the assumption that our WDs are a co-moving, local sample.  After we correct each $v_{\rm app}$ to the Local Standard of Rest (LSR), only random stellar motions dominate the dynamics of our sample.  We assume, for the purposes of this investigation, that these average out.  Thus the mean apparent velocity equals the mean gravitational redshift: $\langle v_{\rm app}\rangle=\langle v_{\rm g}\rangle$.  The idea of averaging over a group of WDs to extract a mean gravitational redshift is not new \citep{greenstein67}, but the availability of an excellent dataset prompted its exploitation.  We address the validity of the co-moving approximation in Section \ref{co-move}.

\section{Observations}\label{obs}

We use spectroscopic data from the European Southern Observatory (ESO) \textbf{S}N Ia \textbf{P}rogenitor surve\textbf{Y} \citep[SPY;][]{napiwotzki01an}.  These observations, taken using the UV-Visual Echelle Spectrograph \citep[UVES;][]{dekker00} at Kueyen, Unit Telescope 2 of the ESO VLT array, constitute the largest, homogeneous, high-resolution (0.36\,\AA\ or $\sim16$\,km s$^{-1}$ at H$\alpha$) spectroscopic dataset for WDs.  We obtain the pipeline-reduced data online through the publicly available ESO Science Archive Facility.

\subsection{Sample}

As explained in \citet{napiwotzki01an}, targets for the SPY sample come from: the white dwarf catalog of \citet{mccook99}, the Hamburg ESO Survey \citep[HES;][]{wisotzki00,christlieb01}, the Hamburg Quasar Survey \citep{hagen95,homeier98}, the Montreal-Cambridge-Tololo survey \citep[MCT;][]{lamontagne00}, and the Edinburgh-Cape survey \citep[EC;][]{kilkenny97}.  The magnitude of the targets is limited to $B<16.5$.

Our main sample consists of 449 analyzed hydrogen-dominated WDs (see Figure \ref{galplot} for the distribution of targets in Galactic coordinates).  This is the subset of the SPY sample that meets our sample criteria (explained below) and that shows measurable $v_{\rm app}$ in the H$\alpha$ (and H$\beta$) line cores while not showing measurable $v_{\rm app}$ variations.  A variable velocity across multiple epochs of observation suggests binarity.  The method of SPY to search for double degenerate systems is to detect variable radial velocity.  For our study, however, we are interested only in non-binary WDs since these presumably have no radial velocity component in addition to random stellar motion after being corrected to the LSR.  We exclude known double degenerates and common proper motion binary systems \citep{finley97b,jordan98,maxted99,maxted00b,silvestri01,koester09b} even if we do not find them to show variable $v_{\rm app}$.

We choose ``normal'' DAs (criterion 1) from \citet{koester09b}.  Classification as a normal DA does not include WDs that exhibit He absorption in their spectra in addition to H absorption, and it does not include magnetic WDs.  In a subsequent paper we will investigate the sample of 20 helium-dominated WDs for which we observe H absorption.

For our main sample, we are also only interested in thin disk WDs (criterion 2), so we exclude halo and thick disk candidates as kinematically classified by \citet{pauli06} and \citet{richter07}.  We assume the rest are thin disk objects, the most numerous Galactic component.  Our sample selection is also consistent with the results for the targets in common with \citet{sion09}.  \citet{richter07} find only 2\% and 6\% of their 632 DA WDs from SPY to be from the halo and thick disk, respectively.  For WDs within 20\,pc, \citet{sion09} find no evidence for halo objects and virtually no thick disk objects.  We note that unique identification of population membership for WDs is difficult and often not possible because of ambiguous kinematical properties.  Based on corrections for these intrinsic contaminations by \citet{napiwotzki09}, we expect any residual contamination in our sample to be at most $\sim6$\%.  A contamination this size will have a negligible impact on our conclusions.  We explain the significance of requiring thin disk WDs in Section \ref{co-move}, and we explore a mini-sample of thick disk WDs in Section \ref{thick}.

The gravitational redshift method becomes very difficult for hot DAs with $50000$\,K $\gtrsim T_{\rm eff}\gtrsim40000$\,K (see the $T_{\rm eff}$ gap in Figure \ref{tdist}).  As the WD cools through this $T_{\rm eff}$ range, the Balmer line core, which we use to measure $v_{\rm app}$ (Section \ref{velocity}), disappears as it transitions from emission to absorption; fortunately only $\sim5\%$ of the DAs from SPY lie in this range.

\begin{figure}[t]
\centering{\includegraphics[width=\columnwidth]{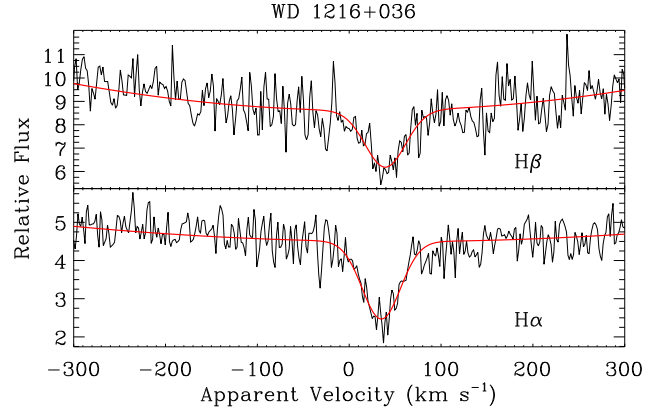}}
\caption{Example UVES spectrum for a target in our sample.  We measure $v_{\rm app}$ by fitting Gaussian profiles ({\it solid, red lines}) to the non-LTE Balmer line cores using a non-linear least-squares fitting routine.  The line cores are well-resolved, allowing for precise centroid determinations.\label{explot}}
\end{figure}

\section{Velocity Measurements}\label{velocity}

In the wings of absorption lines, and in particular, for the hydrogen Balmer series, the effects of collisional broadening cause asymmetry, making it difficult to measure a velocity centroid \citep{shipman76,grabowski87}.  These effects are much less significant, however, in the sharp, non-LTE line cores, and furthermore with decreasing principal quantum number, making both the H$\alpha$ and H$\beta$ line cores suitable options for measuring an apparent velocity $v_{\rm app}$.  Higher order Balmer lines are intrinsically weaker (the H$\gamma$ line core, for example, is seldom observable in our data), so finite S/N prevents the number of observable H$\beta$ line cores from matching the number of observable H$\alpha$ line cores.

We measure $v_{\rm app}$ for each target in our sample by fitting a Gaussian profile to the H$\alpha$ line core using GAUSSFIT, a non-linear least-squares fitting routine in IDL (see Figure \ref{explot} for an example).  When available, we combine this measurement with that of the H$\beta$ line core centroid as a mean weighted according to the uncertainties returned by the fitting routine.  We include H$\beta$ line core centroid information in 372 of our 449 $v_{\rm app}$ measurements.  If multiple epochs of observations exist, we combine these measurements as a weighted mean as well.  Apparent velocity measurements of a given observation (i.e., H$\alpha$ and H$\beta$ line core centroids) are combined before multiple epochs.

Table \ref{table_da} (full version available on-line) shows our measured $v_{\rm app}$ for H$\alpha$ and H$\beta$ (when observed) for each observation.

\begin{figure}
\centering{\includegraphics[width=\columnwidth]{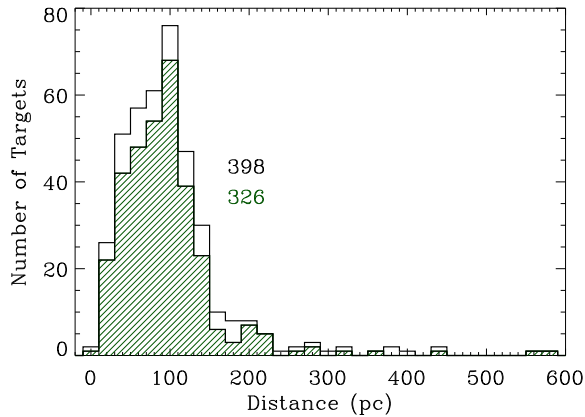}}
\caption{Distribution of distances (from spectroscopic parallax) of SPY WDs from \citet{pauli06}.  The shaded, green histogram shows the targets in our sample.  The mean is 94.5\,pc; the median is 89.2\,pc.  These distances are short enough to support our co-moving approximation.  We list the number of targets in each distribution.\label{ddist}}
\end{figure}

\subsection{Co-Moving Approximation}\label{co-move}

We measure a mean gravitational redshift by assuming that our WDs are a co-moving, local sample.  With this assumption, only random stellar motions dominate the dynamics of our targets; this falls out when we average over the sample.

For this assumption to be valid, at least as an approximation, our WDs must belong to the same kinematic population; in the case of this work, this is the thin disk.  We achieve a co-moving group by correcting each measured $v_{\rm app}$ to the kinematical LSR described by Standard Solar Motion \citep{kerr86}.

There are reasons to believe that the targets in our sample will {\it not} significantly lag behind our choice of LSR due to asymmetric drift.  Although WDs are considered ``old'' since they are evolved stars, it is the total age of the star (main sequence lifetime $\tau_{\rm nuc}$ and cooling time $\tau_{\rm cool}$) that is of consequence.  WDs with $M\sim0.6$\,M$_\odot$ have main sequence progenitors with $M\sim2$\,M$_\odot$ \citep[e.g.,][]{williams09}.  This corresponds to $\tau_{\rm nuc}\sim1.4$\,Gyr \citep{girardi00}.  $\tau_{\rm cool}$ is on the order of a few hundred million years for most of the WDs in our sample ($T_{\rm eff}$ of a few times $10^4$\,K) and $\sim2.5\times10^9$\,yr for our coolest WDs ($T_{\rm eff}\sim7000$\,K); the total age spans a range of roughly 1.5 to 4 Gyr (F/G type stars).

We also make certain that our WDs reside at distances that are small when compared to the size of the Galaxy, thereby making systematics introduced by the Galactic kinematic structure negligible.  Figure \ref{ddist} shows the distances \citep[from spectroscopic parallax;][]{pauli06} to the targets in our sample.  The mean distance of the targets in the histogram is less than 100\,pc, and all are within 600\,pc.  Over these distances, the velocity dispersion with varying height above the disk remains modest \citep{kuijken89}, and differential Galactic rotation changes very little as well \citep[$\sim3$\,km s$^{-1}$;][]{fich89}.  In Section \ref{dynamical}, we perform an empirical check to the assumptions made in this section.

\section{Results}

\begin{figure}
\centering{\includegraphics[width=\columnwidth]{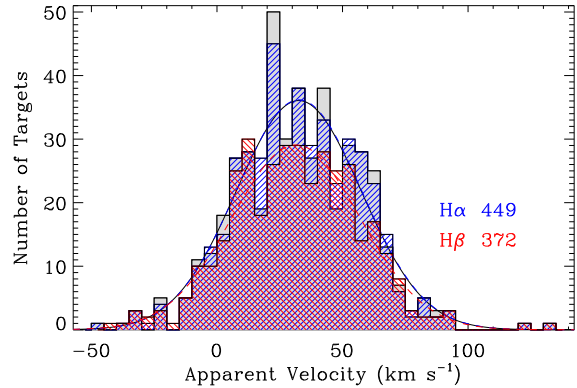}}
\caption{Histograms of measured apparent velocities $v_{\rm app}$ with a bin size of 5\,km~s$^{-1}$.  The mean $v_{\rm app}$ for all targets in our sample ({\it shaded}) is $32.57\pm1.17$\,km s$^{-1}$; the median is $31.94$\,km s$^{-1}$; the standard deviation is $24.84$\,km s$^{-1}$.  Using $v_{\rm app}$ measured from H$\alpha$ only ({\it red, descending lines}): the mean $v_{\rm app}$ is $32.69\pm1.18$\,km s$^{-1}$; the median is $32.05$\,km s$^{-1}$; the standard deviation is $24.87$\,km s$^{-1}$.  Using $v_{\rm app}$ measured from H$\beta$ only ({\it blue, ascending lines}): the mean $v_{\rm app}$ is $31.47\pm1.32$\,km s$^{-1}$; the median is $31.55$\,km s$^{-1}$; the standard deviation is $25.52$\,km s$^{-1}$.  The overplotted curves are the Gaussian distribution functions used to determine Monte Carlo uncertainties.  We list the number of targets in each distribution.\label{vdist}}
\end{figure}

\subsection{Mean Apparent Velocities}\label{vel_text}

We present the distribution of our measured apparent velocities in Figure \ref{vdist}.  Table \ref{table_da} lists individual apparent velocity measurements, and mean apparent velocities are in Table \ref{results}. 

Though our main method uses information from both the H$\alpha$ (Column 7 of Table \ref{table_da}) and H$\beta$ (Column 9) line cores to determine $v_{\rm app}$ for a given observation (Column 11), we also perform our analysis using H$\alpha$ only and H$\beta$ only.  We measure H$\beta$ line core centroids for 382 of our 449 targets.

\begin{figure}[b]
\centering{\includegraphics[width=\columnwidth]{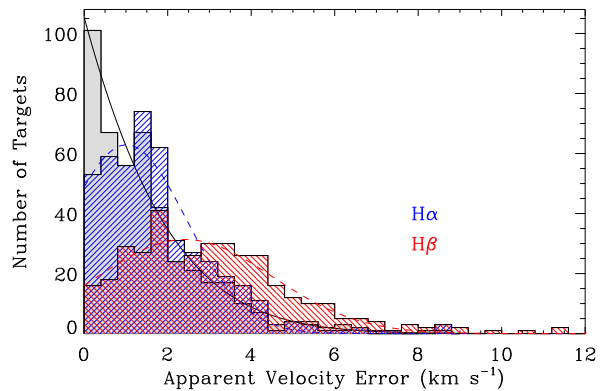}}
\caption{Histograms of apparent velocity measurement uncertainties $\delta v_{\rm app}$ corresponding to the samples in Figure \ref{vdist}.  The bin size is 0.4\,km~s$^{-1}$.  The overplotted curves are the empirical distribution functions used to determine Monte Carlo uncertainties.  Notice that measurements of the H$\alpha$ line core are more precise than for H$\beta$.  For aesthetics, we leave off two H$\beta$ $\delta v_{\rm app}$ of 13.06 and 17.57\,km s$^{-1}$.\label{vdiste}}
\end{figure}

Figure \ref{vdiste} shows the distribution of measurement uncertainties associated with each target.  H$\beta$ centroid determinations are typically less precise than those for H$\alpha$ (see Column 6 of Table \ref{results}), which is expected since the H$\alpha$ line core is nearly always better-defined.  We find that the improved precision achieved by combining H$\alpha$ and H$\beta$ information is not significant when determining the uncertainties to our mean apparent velocities.  These uncertainties are dominated by sample size.  In fact, we must increase (worsen) our typical measurement error of $\sim2$\,km s$^{-1}$ to $\sim10$\,km s$^{-1}$ to notice a $\sim7\%$ increase in the error of the mean; a monstrous leap to measurement uncertainties of $\sim50$\,km s$^{-1}$ enlarges the error of the mean by a little more than a factor of 2.  Thus, using H$\alpha$ (or H$\beta$) centroids only is sufficient for the kind of investigation employed in this paper, and lower resolution observations are also suitable as long as the Balmer line core is resolved.

The quoted uncertainties of the mean apparent velocities (Column 4 of Table \ref{results}) come from Monte Carlo simulations.  For each sample, we recreate a large number of instances (10000) of the $v_{\rm app}$ distribution by randomly sampling from a convolution of the empirical $v_{\rm app}$ distribution (Gaussian characterized by the parameters in Columns 3 and 5 of Table \ref{results}) and the empirical measurement error distribution.  We adopt the standard deviation of the resulting simulated mean values as our formal uncertainties.  Since the input distributions for our simulations are empirical, our uncertainties are subject to the normal limitations of Frequentist statistics.  We plot the empirical distribution of our main sample in Figure \ref{vdist} ({\it black curve}) along with the distributions for the H$\alpha$ ({\it dashed, blue curve}) and H$\beta$ ({\it dashed, red curve}) samples.  The corresponding empirical distributions of our measurement uncertainties are in Figure \ref{vdiste}.

For convenience, Table \ref{results} also lists the quantity $\langle M/R\rangle$, which is proportional to $\langle v_{\rm g}\rangle$ (equation \ref{v_g}) and, as we argue in Section \ref{co-move}, $\langle v_{\rm app}\rangle$.

\subsection{Mean Masses}\label{masses}

The mean apparent velocity $\langle v_{\rm app}\rangle$ (or $\langle M/R\rangle$) is our fundamental result since it is this quantity that is model-independent.  To translate this to a mean mass (Table \ref{table_masses}), we must invoke two dependencies: (1) we need an evolutionary model to give us a mass-radius relation, and (2) since the WD radius does slightly contract during its cooling sequence, we need an estimate of the position along this track for the average WD in our sample (i.e., a mean $T_{\rm eff}$).

\begin{figure}
\centering{\includegraphics[width=\columnwidth]{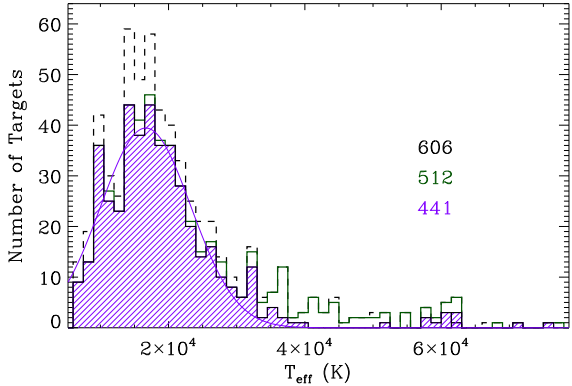}}
\caption{Distribution of spectroscopically determined $T_{\rm eff}$ of normal DAs from \citet{koester09b} ({\it dashed, black histogram}).  The bin size is 1500\,K.  The solid, green histogram shows the non-binary thin disk SPY targets, and the shaded, purple histogram shows the targets in our sample.  The mean is $19400\pm300$\,K; the median is 17611\,K.; the standard deviation is 9950\,K.  The overplotted curve is the empirical distribution function used to determine Monte Carlo uncertainties.  We list the number of targets in each distribution.\label{tdist}}
\end{figure}

\begin{figure*}[p]
\centering{\includegraphics[width=0.80\textwidth]{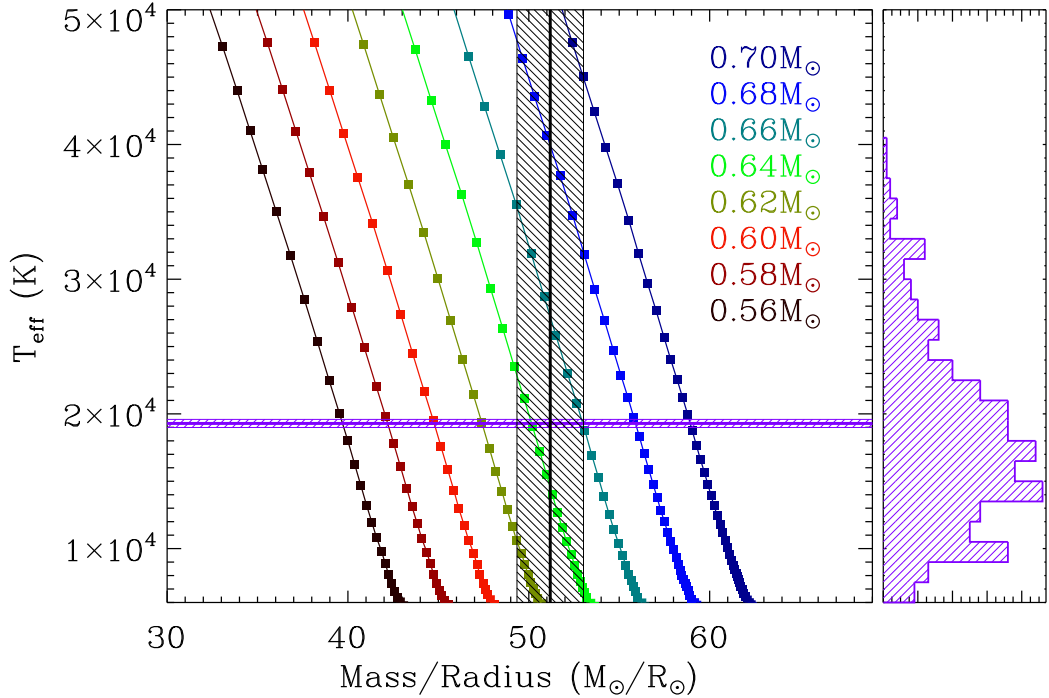}}
\caption{({\it left}) Plot of $M/R$ versus $T_{\rm eff}$ with cooling tracks from evolutionary models for a range of WD masses.  The intersection of the mean measured apparent velocity $v_{\rm app}$ ({\it vertical, black line}) and mean $T_{\rm eff}$ from Figure \ref{tdist} ({\it horizontal, purple line}) indicates a mean mass of $0.647^{+0.013}_{+0.014}$\,M$_\odot$.\label{mrmike2}  ({\it right}) A version of Figure~\ref{tdist} with an abbreviated temperature range.  We leave off 13 WDs with $T_{\rm eff}>50000$\,K from the plot.}
\end{figure*}

Our evolutionary models use $M_{\rm He}/M_\star=10^{-2}$ and $M_{\rm H}/M_\star=10^{-4}$ for the surface-layer masses; these are canonical values derived from evolutionary studies \citep[e.g.,][]{lawlor06}.  See \citet{montgomery99} for a more complete description of our models.  Our dependency on evolutionary models is small.  We are interested in the mass-radius relation from these models, and this is relatively straight-forward since WDs are mainly supported by electron degeneracy pressure, making the WD radius a weak function of temperature. We estimate that varying the C/O ratio in the core affects the radius by less than 0.5\%, whereas changing $M_{\rm H}/M_\star$ from $10^{-4}$ to $10^{-8}$ results in about a 4\% decrease in radius.  See Section \ref{evo} for more discussion on the dependency of the hydrogen layer mass.

Figure \ref{mrmike2} plots $M/R$ versus $T_{\rm eff}$ with cooling tracks from evolutionary models for a range of WD masses.  We use $\langle T_{\rm eff}\rangle=19400\pm300$\,K from the spectroscopically determined values of \citet{koester09b} (see Figure \ref{tdist}), and, after plotting $\langle M/R\rangle$ from Table \ref{results}, we interpolate to arrive at a mean mass of $0.647^{+0.013}_{-0.014}$\,M$_\odot$ for 449 non-binary thin disk normal DA WDs from the SPY sample.

\begin{figure*}[p]
\centering{\includegraphics[width=0.80\textwidth]{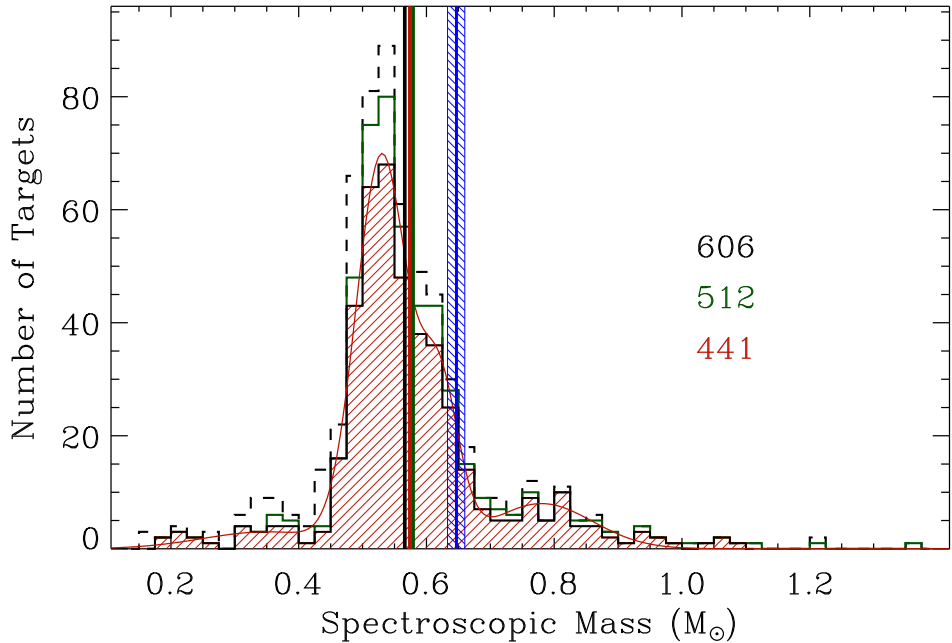}}
\caption{Distribution of spectroscopic masses of normal DAs from \citet{koester09b} we derive using the published atmospheric parameters log\,$g$ and $T_{\rm eff}$ ({\it dashed, black histogram}).  The bin size is 0.025\,M$_\odot$.  The solid, green histogram shows the non-binary thin disk SPY targets, and the shaded, orange histogram shows the targets in our sample.  The means are $0.567\pm0.002$\,M$_\odot$ ({\it vertical, black line}), $0.580\pm0.002$\,M$_\odot$ ({\it vertical, green line}), and $0.575\pm0.002$\,M$_\odot$ ({\it vertical, orange line}), respectively.  Notice that the mean spectroscopic masses are similar, indicating that the application of our sample criteria to SPY is not introducing additional systematic effects.  All the means are also significantly less than the mean mass derived from the gravitational redshift method ({\it vertical, blue line}).  The overplotted curve is the empirical distribution function used to determine Monte Carlo uncertainties.  We list the number of targets in each distribution.\label{massdist}}
\end{figure*}

To compare this result with that of the spectroscopic method, we use atmospheric parameters log\,$g$ and $T_{\rm eff}$ from \citet{koester09b} along with the mass-radius relation from evolutionary models to derive individual masses for 441 of the targets in our sample (\citet{koester09b} do not publish individual WD masses).  We derive a sharply peaked mass distribution (Figure \ref{massdist}) with width (not uncertainty) $\sigma=0.13$\,M$_\odot$ and a mean mass of $0.575\pm0.002$\,M$_\odot$ $-$ significantly lower than the value we obtain from the gravitational redshift method.  We compute the error of the mean using Monte Carlo simulations following the same method described in Section \ref{vel_text} except instead of using a single Gaussian to represent the mass distribution, we use multiple Gaussians (curve in Figure \ref{massdist}).

\subsection{Systematic Effects}\label{systematics}

\subsubsection{From Evolutionary Models}\label{evo}

The hydrogen layer mass in DAs is believed to be in the range of $10^{-4}\gtrsim M_{\rm H}/M_\star\gtrsim 10^{-8}$, constrained by hydrogen shell burning in the late stages of stellar evolution \citep{althaus02,lawlor06} and convective mixing \citep{fontaine97}.  In their asteroseismological studies, \citet{bischoff-kim08} also find evidence to support this range of hydrogen layer masses, and this is consistent with the results of \citet{castanheira09}.

Our evolutionary models use the fiducial value of $M_{\rm H}/M_\star=10^{-4}$ for ``thick'' hydrogen layers. First, this is suggested by the pre-white dwarf evolutionary models of, e.g., \citet{lawlor06}, who find that the overwhelming majority of their DA models have thick hydrogen layers. Second, if thin layers were the norm, then convective mixing below 10000\,K would lead to a disappearance of DAs at these temperatures \citep{fontaine97}. Both of these reasons lead us to choose thick hydrogen layers for our models. 

We find that using a midrange hydrogen layer mass of $M_{\rm H}/M_\star=10^{-6}$ decreases the mean mass we derive for our main sample by $0.012$\,M$_\odot$, while using a thin layer mass of $M_{\rm H}/M_\star=10^{-8}$ decreases the derived mean mass by an additional $0.003$\,M$_\odot$ (total mass difference of $0.015$\,M$_\odot$).  Assuming no hydrogen layer ($M_{\rm H}/M_\star=0$) yields a mean mass that is $\sim0.018$\,M$_\odot$ lower than that obtained with the fiducial value of $M_{\rm H}/M_\star=10^{-4}$.

It is worth noting that the spectroscopic method shares this dependency on evolutionary models and that most of the studies listed in Table \ref{previous}, including \citet{liebert05}, \citet{kepler07} and \citet{tremblay09}, employ mass-radius relations that use thick hydrogen layers.  Column 7 of Table \ref{previous} notes the assumed hydrogen layer mass in the evolutionary models used in each study.  Furthermore, our results are qualitatively less sensitive to the mass-radius relation: for the gravitational redshift method, $v_{\rm g}\propto M/R$, while the surface gravity used by the spectroscopic method scales as $g\propto M/R^2$.

\subsubsection{Dynamical}\label{dynamical}

We use the kinematical LSR described by Standard Solar Motion \citep{kerr86} as our reference frame for the co-moving approximation.  To determine if this is a suitable choice, we investigate $\langle v_{\rm app}\rangle$ in the $U$, $V$ or $W$ directions (by convention, $U$ is positive toward the Galactic center, $V$ is positive in the direction of Galactic rotation, and $W$ is positive toward the North Galactic Pole).

For 237 targets in the direction of the Galactic center ($l\le90^\circ$ or $l\ge270^\circ$) and 212 opposite the Galactic center (90$^\circ<l<270^\circ$), $\langle v_{\rm app}\rangle=31.81\pm1.71$ and $33.43\pm1.64$\,km s$^{-1}$, respectively.  In the direction of the LSR flow ($l=90^\circ,b=0^\circ$; 196 targets) and opposite the flow (253 targets), $\langle v_{\rm app}\rangle=33.61\pm2.09$ and $31.77\pm1.34$\,km s$^{-1}$.  North (185) and south (264) of the Galactic equator, $\langle v_{\rm app}\rangle=31.59\pm1.84$ and $33.26\pm1.53$\,km s$^{-1}$.

These empirical checks provide independent evidence that the local WDs in our sample move with respect to kinematical LSR with the following values: ($U,V,W$)=($-1.62\pm3.35,+1.84\pm3.43,-1.67\pm3.37$)\,km s$^{-1}$, which is consistent with {\it no} movement relative to the LSR.  Therefore, we find our choice of reference frame to be suitable for this study.

\subsubsection{Observational}

SPY targets are magnitude-limited to $B<16.5$, but these targets come from multiple surveys with varying selection criteria, making the combined criteria difficult to precisely determine \citep{koester09b}.  For this reason, our results pertain mostly to non-binary thin disk normal DA WDs from SPY.  Although the selection bias is likely to have a minimal effect, a detailed comparison of our results with that of the general DA population awaits a closer examination of the selection criteria \citep[see][]{napiwotzki01an,napiwotzki03}.

If we approximate our sample to be free of any target selection bias, our crude estimates show that we have a net observational bias toward lower mass WDs.  There are two competing effects: first, at a given $T_{\rm eff}$, a larger mass (smaller radius) results in a fainter WD, thus biasing the detection of fewer higher mass WDs over a given volume, and second, a larger mass (smaller radius) also results in a slower cooling rate due to a larger heat capacity as well as a diminished surface area.  This means more higher mass WDs as a function of $T_{\rm eff}$.  We estimate the observational mass bias correction as follows:

Let $P(M)$ be the distribution of WDs as a function of mass for a magnitude-limited sample of WDs. For simplicity, we take it to have the form of a Gaussian; we take the mean to be $\langle M \rangle \sim0.65$\,M$_\odot$ and $\sigma\sim0.1$\,M$_\odot$.  As a reference, the spectroscopic mass distribution of DAs shows a sharp Gaussian-like peak with high and low mass wings \citep[e.g.,][]{bergeron92b,liebert05,kepler07}.

Effect (1): Ignoring color, the apparent flux of a star scales as
$F_{\rm app}\sim L_{\star}/D^2$ and the luminosity as $L_{\star}\sim
R^2T_{\rm eff}^4$, where $L_\star$, $R$ and $T_{\rm eff}$ are the luminosity, radius, and effective temperature of the star; $D$ is its distance.  In the non-relativistic limit, the radius $R$ of a WD scales as $R\propto M^{-1/3}$ \citep{chandrasekhar39}, and for a (moderately relativistic) 0.6\,M$_\odot$ WD this relation is approximately $R\propto M^{-1/2}$, so
\begin{equation}
L_\star\propto\frac{T_{\rm eff}^4}{M}.
\label{lum}
\end{equation}
If $F_{\rm cutoff}$ is the lower limit on flux for the survey, a given WD is visible out to a distance of
\begin{equation}
D\sim\left(\frac{L_\star}{F_{\rm cutoff}}\right)^{1/2}\propto\frac{T_{\rm eff}^2}{M^{1/2}}.
\end{equation}
If we make the simplifying assumption that all the WDs are at the observed average temperature $\langle T_{\rm eff}\rangle$ and that they are distributed uniformly, the volume $V$ in which a WD is visible is
\begin{equation}
V\sim D^3\propto M^{-\frac{3}{2}}.
\end{equation}
Thus, $P(M)$ is biased by this factor.

Effect (2): From simple Mestel theory \citep{mestel52}, the WD cooling time $\tau$ scales as
\begin{equation}
\tau\propto\left(\frac{M}{L_\star}\right)^\frac{5}{7},
\end{equation}
which, from equation~\ref{lum}, yields
\begin{equation}
\tau\propto \left(\frac{M^2}{T_{\rm eff}^4}\right)^\frac{5}{7} \sim M^{10/7} T_{\rm eff}^{-20/7}.
\end{equation}
Again, assuming that the WDs are all at $\langle T_{\rm eff}\rangle$, the observed distribution will be biased by a factor of $\tau \propto M^{10/7}$.

Thus, the final biased distribution we observe is given by the product of these factors:
\begin{eqnarray}
P_{\rm bias}(M) & \propto & V \,\tau \, P(M)\nonumber \\
  & \propto & M^{-1/14} P(M).
\end{eqnarray}
This very weak mass bias results in $\langle M \rangle_{\rm bias}=0.649$\,$M_\odot$, which is a mass bias of $\Delta M=-0.001$\,$M_\odot$.  While this is just a crude estimate, it suggests that the bias correction is likely much smaller than the size of our stated random uncertainties.

\subsubsection{Mass Conversion}

In our mean mass determination in Section \ref{masses}, we implicitly assume that $\langle M/R \rangle = \langle M \rangle / \langle R \rangle $.  These quantities are not entirely equal, and by performing an estimate using a simple analytical form for the WD mass distribution, we find that there is a difference of $\sim0.5\%$ (i.e., $\langle M/R \rangle \simeq 1.005 \times \langle M \rangle / \langle R \rangle $), which we consider to be a negligible systematic.

\begin{figure}[t]
\centering{\includegraphics[width=\columnwidth]{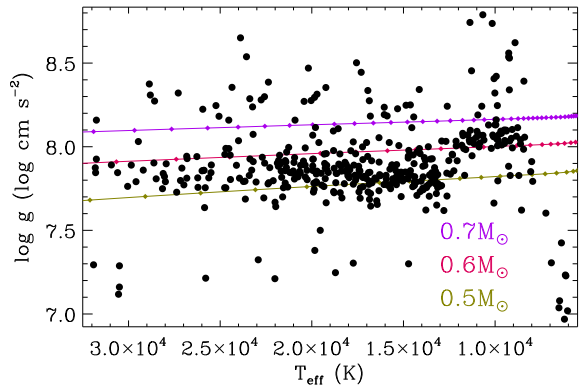}}
\caption{Distribution of $T_{\rm eff}$ versus log\,$g$ for 419 of our WDs.  Spectroscopic parameters for all targets are from \citet{koester09b}.  Notice the abrupt increase in the mean log\,$g$ around 12000\,K.  We also plot cooling tracks from evolutionary models for 0.5, 0.6 and 0.7\,M$_\odot$ WDs.\label{upturn}}
\end{figure}

\subsection{Thick Disk DAs}\label{thick}

The kinematics of thick disk stars prohibit us from placing them in the same co-moving reference frame as thin disk stars.  In Section \ref{dynamical}, we show that the kinematical LSR described by Standard Solar Motion is a suitable choice of reference frame for the SPY {\it thin disk} WDs.  As expected, using $v_{\rm app}$ of our thick disk targets corrected to that LSR (the reference frame suitable for the thin disk) give discrepant values for $\langle v_{\rm app}\rangle$ in opposite directions.  Since our thick disk sample is small (26 targets), our $\langle v_{\rm app}\rangle$ uncertainties are too large to discern a suitable reference frame.  If we correct by the average lag in rotational velocity of the thick disk with respect to the thin disk \citep[$\sim40$\,km s$^{-1}$;][]{gilmore89}, then $\langle v_{\rm app}\rangle=32.90\pm9.59$\,km s$^{-1}$ for our thick disk sample.  Individual $v_{\rm app}$ measurements are listed in Table \ref{table_thick}.  Using $\langle T_{\rm eff}\rangle=19960$\,K, we find $\langle M\rangle=0.652^{+0.097}_{-0.119}$\,M$_\odot$, which is evidence that the mean mass of thick disk DAs is the same as for thin disk DAs.

One should also notice that the dispersion of $v_{\rm app}$ (Column 5 of Table \ref{results}) is clearly larger than that for the thin disk DAs.  Since the $v_{\rm app}$ distribution is a convolution of the true mass distribution and the random stellar velocity distribution, this is consistent with a larger velocity dispersion as expected for the thick disk population \citep{gilmore89}.

\section{Discussion}

\subsection{The Log\,$g$ Upturn}\label{upturn_text}

\begin{figure}[t]
\centering{\includegraphics[width=\columnwidth]{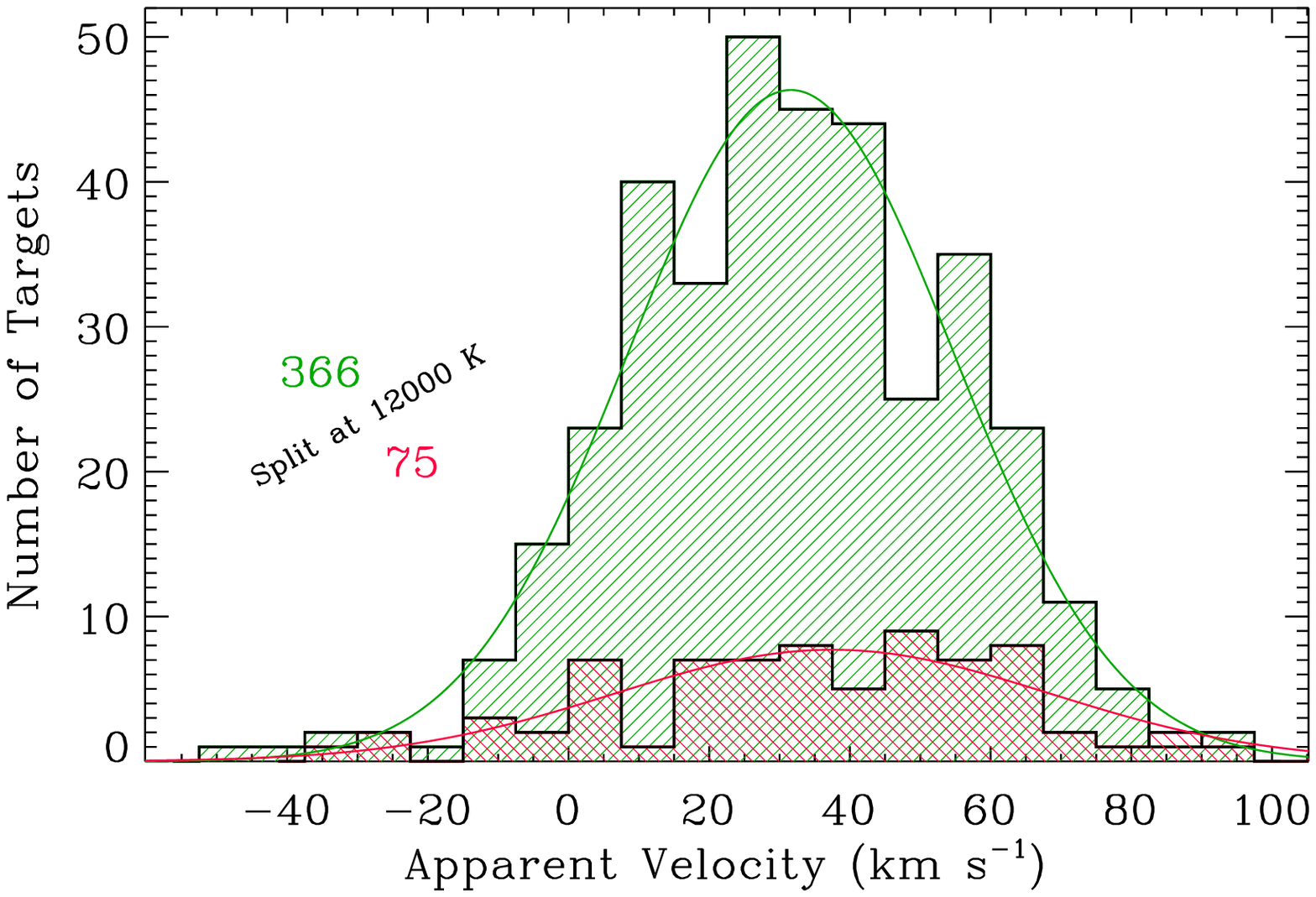}}
\caption{Histogram of measured apparent velocities $v_{\rm app}$ for targets with spectroscopically determined $T_{\rm eff}$ from \citet{koester09b}.  The bin size is 7.5\,km~s$^{-1}$.  The green histogram with ascending lines corresponds to targets with $T_{\rm eff}>12000$\,K and the pink histogram with descending lines to $12000$\,K\,$>T_{\rm eff}>7000$\,K.  The mean $v_{\rm app}$ for the green histogram is $31.61\pm1.22$\,km s$^{-1}$; the median is $31.71$\,km s$^{-1}$; the standard deviation is $23.22$\,km s$^{-1}$.  The mean $v_{\rm app}$ for the pink histogram is $37.50\pm3.59$\,km s$^{-1}$; the median is $36.20$\,km s$^{-1}$; the standard deviation is $31.00$\,km s$^{-1}$.  The overplotted curves are the Gaussian distribution functions used to determine Monte Carlo uncertainties.  We list the number of targets in each distribution.\label{tsplit}}
\end{figure}

\subsubsection{The Problem}

A major problem plaguing the field of WDs is the apparent systematic increase in mean log\,$g$ for DAs toward low ($\lesssim12,000$\,K) $T_{\rm eff}$, as determined from spectroscopic fitting of absorption line profiles \citep{bergeron07,koester09a}.  This increase is absent in photometric log\,$g$ determinations \citep{kepler07,engelbrecht07}, which are not strongly dependent on line profiles.  A number of effects are known to exist that make theoretical line profile modeling for cool WD atmospheres more difficult than for hotter WDs, such as helium contamination from dredge-up \citep{bergeron90,tremblay08} and the treatment of convective efficiency \citep{bergeron95a}.  Neither of these, however, seem to solve the log\,$g$ upturn problem \citep{koester09a}, and since no strong hypotheses have been put forth to explain a real increase in mean mass \citep{kepler07}, the fault most likely lies with the atmosphere models or with the limitations of these models.

The number of studied cool WDs is already relatively low due to the inherent difficulty of observing cool objects, but the addition of the log\,$g$ upturn problem and the subtleties of cool WD atmosphere modeling has thus far kept that number low by prompting many spectroscopic analyses to be designed to exclude cooler WDs \citep[e.g.,][]{bergeron92b,madej04,liebert05,kepler07}.  This is tremendously unfortunate.  Understanding cool WDs has broad astrophysical relevance, such as in determining the age of the Galactic disk \citep{winget87} and in setting constraints on the physics of crystallization in high-density plasmas \citep{winget09}.

Furthermore, decades of focus on hotter WDs (due to the much larger dataset and due to the neglect of cooler WDs) have perhaps given researchers in our field a false comfort with these objects.  There is a feeling that since hot WD atmospheres are more straightforward to model than cool atmospheres, the spectroscopic surface gravities (and masses) must be correct for the hot WDs and not for the cool WDs, given the log\,$g$ upturn problem.  Recent improved calculations for Stark broadening of hydrogen lines in WD atmospheres \citep{tremblay09} show that hot WD modeling is still maturing.

\subsubsection{Avoiding the Upturn}

The gravitational redshift method is independent of log\,$g$ determinations from atmosphere models and allows us to constrain changes in mean masses across $T_{\rm eff}$ bins.

\begin{figure}
\centering{\includegraphics[width=\columnwidth]{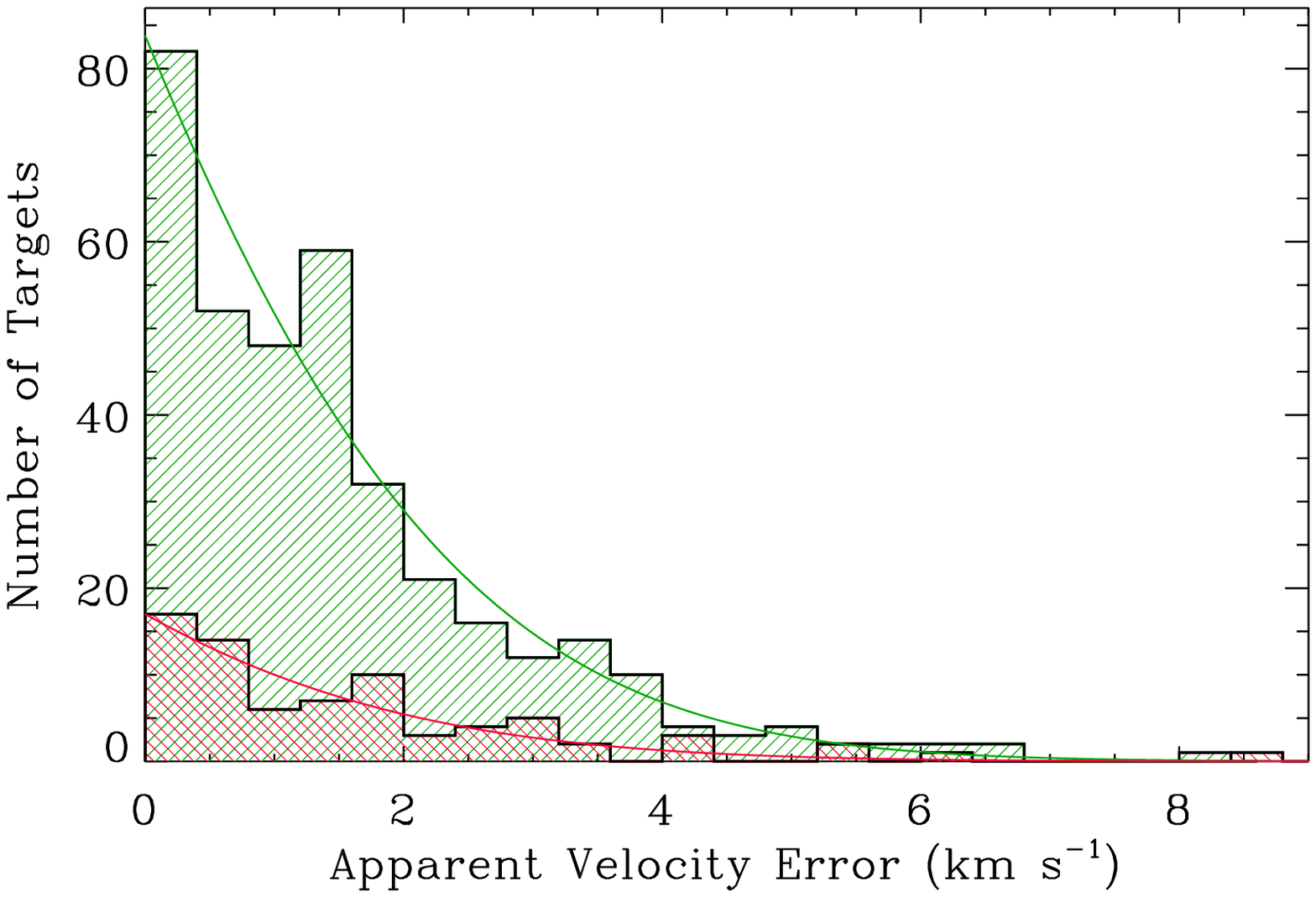}}
\caption{Histogram of measured apparent velocities $v_{\rm app}$ for targets with spectroscopically determined $T_{\rm eff}$ from \citet{koester09b}.  The bin size is 7.5\,km~s$^{-1}$.  The green histogram with ascending lines corresponds to targets with $T_{\rm eff}>12000$\,K and the pink histogram with descending lines to $12000$\,K\,$>T_{\rm eff}>7000$\,K.  The mean $v_{\rm app}$ for the green histogram is $31.61\pm1.22$\,km s$^{-1}$; the median is $31.71$\,km s$^{-1}$; the standard deviation is $23.22$\,km s$^{-1}$.  The mean $v_{\rm app}$ for the pink histogram is $37.50\pm3.59$\,km s$^{-1}$; the median is $36.20$\,km s$^{-1}$; the standard deviation is $31.00$\,km s$^{-1}$.  The overplotted curves are the Gaussian distribution functions used to determine Monte Carlo uncertainties.  We list the number of targets in each distribution.\label{tsplit}}
\caption{Similar to Figure \ref{vdiste} but corresponding to targets with $T_{\rm eff}>12000$\,K ({\it green}) and to targets with $12000$\,K\,$>T_{\rm eff}>7000$\,K ({\it pink}).\label{tsplite}}
\end{figure}

Figure \ref{upturn} plots spectroscopically determined values of log\,$g$ and $T_{\rm eff}$ from \citet{koester09b} for the targets in our sample, clearly exposing the upturn.  We plot evolutionary models for 0.5, 0.6 and 0.7\,M$_\odot$ DA WDs to illustrate how a higher surface gravity implies a higher mass and to show the expected weak dependence on $T_{\rm eff}$.  Using the mass-radius relation from evolutionary models, we derive mean spectroscopic masses $\langle M\rangle$\,$^{\rm hot}=0.563\pm0.002$\,M$_\odot$ for 358 WDs with $T_{\rm eff}>12000$\,K and $\langle M\rangle$\,$^{\rm cool}=0.666\pm0.005$\,M$_\odot$ for 75 WDs with $12000$\,K\,$>T_{\rm eff}>7000$\,K; $\Delta\langle M\rangle=0.103\pm0.007$\,M$_\odot$.  The mass difference is even larger in the SDSS data; \citet{kepler07} find $\langle M\rangle$\,$^{\rm hot}=0.593\pm0.016$\,M$_\odot$ and $\langle M\rangle$\,$^{\rm cool}=0.789\pm0.005$\,M$_\odot$ ($12000$\,K\,$\ge T_{\rm eff}\ge8500$\,K); $\Delta\langle M\rangle=0.196\pm0.021$\,M$_\odot$. 

\begin{figure}
\centering{\includegraphics[width=\columnwidth]{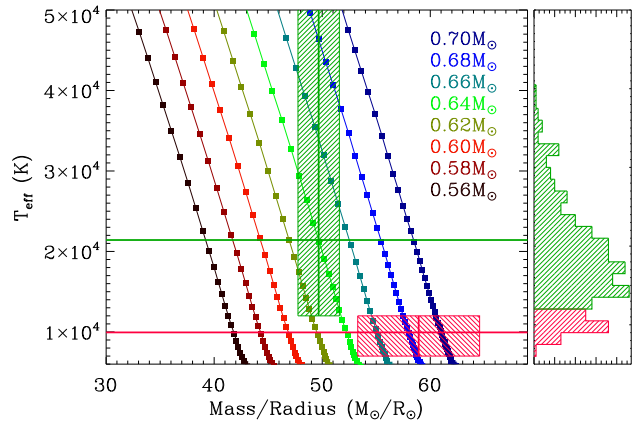}}
\caption{Same as Figure \ref{mrmike2} but for targets with $T_{\rm eff}>12000$\,K ({\it green}) and with $12000$\,K\,$>T_{\rm eff}>7000$\,K ({\it pink}).  $\langle M\rangle$\,$^{\rm hot}=0.640\pm0.014$\,M$_\odot$ and $\langle M\rangle$\,$^{\rm cool}=0.686^{+0.035}_{-0.039}$\,M$_\odot$.\label{mrmikesplit}}
\end{figure}

In Figure \ref{tsplit}, we show our distribution of $v_{\rm app}$ (distribution of uncertainties in Figure \ref{tsplite}) for targets with $T_{\rm eff}>12000$\,K ({\it green histogram with ascending lines}) and with $12000$\,K\,$>T_{\rm eff}>7000$\,K ({\it pink histogram with descending lines}).  The corresponding $\langle v_{\rm app}\rangle$ determinations are $31.61\pm1.22$ and $37.50\pm3.59$\,km s$^{-1}$, respectively, which translates to $\langle M\rangle$\,$^{\rm hot}=0.640\pm0.014$\,M$_\odot$ and $\langle M\rangle$\,$^{\rm cool}=0.686^{+0.035}_{-0.039}$\,M$_\odot$ (see Figure \ref{mrmikesplit}).  This is consistent with {\it no} change in mean mass across a temperature split at $T_{\rm eff}=12000$\,K in agreement with the photometric studies by \citet{kepler07} and by \citet{engelbrecht07}.  No previous large spectroscopic study has seen consistency in mean mass across these temperatures.

\subsection{Comparison With Previous Studies}\label{previoustext}

Table \ref{previous} lists four studies that employ the gravitational redshift method to determine masses for common proper motion WDs.  Because of the small sample sizes (9, 35, 34, and 41 WDs), the uncertainties of the mean masses found by these studies are relatively large $-$ too large to discern a difference in mean mass from that of the spectroscopic method \citep{silvestri01}.  Other than with the results of \citet{koester87}, whose sample consisted of only 9 DAs, our mean mass agrees with that of all these studies, and we improve upon the uncertainties (precision) by more than a factor of two.

The mean mass of 512 SPY non-binary thin disk normal DAs from \citet{koester09b}, as we figure from their spectroscopically determined values of log\,$g$ and $T_{\rm eff}$, is $0.580\pm0.002$\,M$_\odot$, and if we restrict the comparison to 441 WDs in our sample, $\langle M\rangle$\,$=0.575\pm0.002$\,M$_\odot$.  Both values are significantly smaller than the mean mass we derive using the gravitational redshift method.  

Using atmosphere models that implement the new Stark broadened line profiles from \citet{tremblay09} and an updated treatment of the microfield distribution, the SPY sample shows an increase of $\sim0.03$\,M$_\odot$ in the mean mass (Koester, private communication), but this resulting mean mass is still significantly less than our value.  In fact, our mean mass is significantly larger than the determinations from all the previous spectroscopic studies listed in Table \ref{previous} except that of \citet{tremblay09}.  

The recent work of \citet{tremblay09} uses atmosphere models with improved Stark broadening calculations to reanalyze the WDs from \citet{liebert05}.  They find a larger mean mass (0.649\,M$_\odot$) than previously determined for the Palomar-Green sample (0.603\,M$_\odot$).  The mean mass we derive using the gravitational redshift method agrees well, thus providing independent observational evidence in support of these improved atmosphere models.

\section{Conclusions}

We show that the gravitational redshift method can be used to determine a mean mass of a sample of WDs whose dynamics are dominated by random stellar motions.  For 449 non-binary thin disk normal DA WDs from SPY, we find $\langle v_{\rm g}\rangle=\langle v_{\rm app}\rangle=32.57\pm1.16$\,km s$^{-1}$.  Using the mass-radius relation from evolutionary models, $\langle M\rangle=0.647^{+0.013}_{-0.014}$\,M$_\odot$.  This is in agreement with the results of previous gravitational redshift studies, but it is significantly higher than all previous spectroscopic determinations except that of \citet{tremblay09}.

We find that the targets in our sample move with respect to the kinematical LSR described by Standard Solar Motion \citep{kerr86} with the following values: ($U,V,W$)=($-1.62\pm3.35,+1.84\pm3.43,-1.67\pm3.37$)\,km s$^{-1}$.  {\it This is consistent with no movement relative to this LSR.}

Our results provide evidence that the mean mass of thick disk DAs is the same as for thin disk DAs.

The gravitational redshift method is independent of spectroscopically determined surface gravity from atmosphere models and is insensitive to the log\,$g$ upturn problem (Section \ref{upturn_text}).  $\langle v_{\rm app}\rangle=31.61\pm1.22$ and $37.50\pm3.59$\,km s$^{-1}$ for targets with $T_{\rm eff}>12000$\,K and with $12000$\,K\,$>T_{\rm eff}>7000$\,K, respectively.  This translates to $\langle M\rangle$\,$^{\rm hot}=0.640\pm0.014$\,M$_\odot$ and $\langle M\rangle$\,$^{\rm cool}=0.686^{+0.035}_{-0.039}$\,M$_\odot$, which disagrees with spectroscopic results by showing {\it no} significant change in the mean mass of DAs across a temperature split at $T_{\rm eff}=12000$\,K.  This confirms the resuls of \citet{kepler07} and \citet{engelbrecht07}, who find no log\,$g$ increase in their photometric investigations.  We are currently obtaining more observations of cool WDs to increase our sample size and hence precision of our mean mass determinations.

\acknowledgements 
R.E.F. would like thank the SPY collaboration for providing spectra of excellent quality.  The observations were made with the European Southern Observatory telescopes and obtained from the ESO/ST-ECF Science Archive Facility.  Thank you, Detlev Koester, for supplying an early version of the final DA SPY paper.  Thank you, Seth Redfield, for guidance at the start of this project.  We also thank Kepler Oliveira and the referee for helpful comments.  This work has made use of NASA's Astrophysics Data System Bibliographic Services. It has also made use of the SIMBAD database, operated at CDS, Strasbourg, France.  This work is supported by the National Science Foundation under grant AST-0909107, the Norman Hackerman Advanced Research Program under grant 003658-0255-2007, and the Joint Institute for High Energy Density Science, funded by The University of Texas System and supported in part by Sandia National Laboratories. M.H.M acknowledges the support of the Delaware Asteroseismic Center.

\bibliographystyle{apj}
\bibliography{/home/grad79/cylver/all}


\clearpage

\begin{landscape}
\begin{deluxetable*}{cccccccccccccc}
\tablewidth{0pt}
\tabletypesize{\scriptsize}
\tablecaption{Apparent Velocity Measurements for Normal DA WDs (Abbreviated)\label{table_da}}
\tablehead{
\colhead{} & \multicolumn{2}{c}{Adopted} & \colhead{} & \colhead{} & \colhead{LSR} & \multicolumn{2}{c}{H$\alpha$} & \colhead{} & \multicolumn{2}{c}{H$\beta$} & \colhead{} & \multicolumn{2}{c}{Observation}
\\
\cline{2-3} \cline{7-8} \cline{10-11} \cline{13-14} \vspace{-0.12in}
\\
\colhead{} & \colhead{$v_{\rm app}$} & \colhead{$\delta v_{\rm app}$} & \colhead{Date} & \colhead{Time} & \colhead{Correction} & \colhead{$v_{\rm app}$} & \colhead{$\delta v_{\rm app}$} & \colhead{} & \colhead{$v_{\rm app}$} & \colhead{$\delta v_{\rm app}$} & \colhead{} & \colhead{$v_{\rm app}$} & \colhead{$\delta v_{\rm app}$}
\\
\colhead{Target} & \colhead{(km s$^{-1}$)} & \colhead{(km s$^{-1}$)} & \colhead{(UT)} & \colhead{(UT)} & \colhead{(km s$^{-1}$)} & \colhead{(km s$^{-1}$)} & \colhead{(km s$^{-1}$)} & \colhead{} & \colhead{(km s$^{-1}$)} & \colhead{(km s$^{-1}$)} & \colhead{} & \colhead{(km s$^{-1}$)} & \colhead{(km s$^{-1}$)}
}
\startdata
WD 0000-186      & 24.530  & 0.015     & 2000.09.16  & 04:53:07  & -3.511  & 24.515  & 1.267   &   & 24.696  & 4.013   &   & 24.531  & 0.073   \\
                 &         &           & 2000.09.17  & 03:27:31  & -3.825  & 24.190  & 0.742   &   & 29.329  & 4.231   &   & 24.343  & 1.236   \\
HS 0002+1635     & 23.518  & 2.450     & 2002.12.02  & 01:07:24  & -22.829 & 23.518  & 2.450   &   & \nodata & \nodata &   & 23.518  & 2.450   \\
WD 0005-163      & 15.006  & 0.005     & 2000.09.16  & 03:31:59  & -2.013  & 15.057  & 1.892   &   & 14.814  & 3.648   &   & 15.006  & 0.140   \\
                 &         &           & 2002.08.04  & 10:00:19  & 16.515  & 15.921  & 1.860   &   & 9.051   & 4.473   &   & 14.907  & 3.445   \\
WD 0011+000      & 25.655  & 0.106     & 2000.07.14  & 07:14:10  & 28.209  & 23.079  & 0.949   &   & 20.950  & 2.568   &   & 22.823  & 0.978   \\
                 &         &           & 2000.07.17  & 07:38:21  & 27.631  & 25.660  & 0.657   &   & 25.542  & 4.107   &   & 25.657  & 0.025   \\
WD 0013-241      & 15.760  & 0.061     & 2000.09.16  & 02:44:05  & -3.848  & 15.754  & 1.063   &   & 15.797  & 2.591   &   & 15.760  & 0.020   \\
                 &         &           & 2000.09.17  & 01:52:24  & -4.237  & 13.188  & 1.268   &   & 10.630  & 2.367   &   & 12.617  & 1.505   \\
WD 0016-258      & 44.969  & 1.523     & 2000.09.16  & 03:01:00  & -4.332  & 45.801  & 1.451   &   & \nodata & \nodata &   & 45.801  & 1.451   \\
                 &         &           & 2000.09.17  & 02:09:57  & -4.713  & 44.016  & 2.194   &   & 39.586  & 6.581   &   & 43.573  & 1.879   \\
WD 0016-220      & 10.875  & 1.715     & 2000.09.16  & 05:11:37  & -2.989  & 12.101  & 0.868   &   & 16.054  & 1.894   &   & 12.788  & 2.117   \\
                 &         &           & 2000.09.17  & 03:47:05  & -3.294  & 10.506  & 0.742   &   & 7.857   & 1.757   &   & 10.105  & 1.343   \\
WD 0017+061      & -1.247  & 3.824     & 2002.09.26  & 07:34:49  & 2.674   & -0.139  & 2.876   &   & -7.848  & 7.022   &   & -1.247  & 3.824   \\
WD 0018-339      & 30.744  & 0.565     & 2002.09.15  & 02:14:27  & -6.478  & 31.118  & 1.042   &   & 29.443  & 2.256   &   & 30.823  & 0.901   \\
                 &         &           & 2002.09.18  & 02:33:07  & -7.790  & 30.220  & 1.117   &   & 21.817  & 2.406   &   & 28.729  & 4.539   \\
WD 0024-556      & 84.029  & 2.130     & 2000.08.03  & 09:18:35  & -1.148  & 84.420  & 1.490   &   & 78.216  & 5.749   &   & 84.029  & 2.130   \\
WD 0029-181      & 42.792  & 0.136     & 2002.09.26  & 08:24:34  & -4.938  & 40.163  & 1.568   &   & 46.906  & 1.747   &   & 43.170  & 4.740   \\
                 &         &           & 2002.09.27  & 06:13:41  & -5.196  & 42.249  & 1.494   &   & 44.187  & 2.475   &   & 42.767  & 1.212   \\
HE 0031-5525     & 39.229  & 6.206     & 2001.12.17  & 00:53:54  & -25.943 & 37.092  & 1.627   &   & \nodata & \nodata &   & 37.092  & 1.627   \\
                 &         &           & 2002.07.27  & 06:08:08  & 1.501   & 49.230  & 1.699   &   & 42.597  & 4.058   &   & 48.241  & 3.342   \\
HE 0032-2744     & 48.911  & 3.650     & 2002.09.15  & 02:59:55  & -2.771  & 51.692  & 2.379   &   & \nodata & \nodata &   & 51.692  & 2.379   \\
                 &         &           & 2002.09.18  & 03:06:41  & -4.129  & 46.515  & 2.208   &   & \nodata & \nodata &   & 46.515  & 2.208   \\
WD 0032-175      & 31.342  & 1.349     & 2002.09.18  & 03:21:03  & -0.053  & 30.550  & 0.995   &   & \nodata & \nodata &   & 30.550  & 0.995   \\
                 &         &           & 2002.09.25  & 06:31:34  & -3.823  & 32.492  & 1.200   &   & \nodata & \nodata &   & 32.492  & 1.200   \\
WD 0032-177      & 9.875   & 3.692     & 2002.09.18  & 03:29:57  & -0.101  & 10.312  & 1.452   &   & 6.331   & 1.284   &   & 8.078   & 2.793   \\
                 &         &           & 2002.09.25  & 06:01:55  & -3.790  & 15.685  & 1.756   &   & 9.583   & 2.500   &   & 13.668  & 4.059   \\
WD 0033+016      & 88.746  & 2.451     & 2002.09.26  & 07:57:13  & 2.732   & 88.746  & 2.451   &   & \nodata & \nodata &   & 88.746  & 2.451   \\
MCT 0033-3440    & 47.910  & 0.286     & 2000.09.16  & 04:36:44  & -6.179  & 52.870  & 1.751   &   & \nodata & \nodata &   & 52.870  & 1.751   \\
                 &         &           & 2002.08.15  & 09:54:03  & 6.272   & 47.879  & 1.316   &   & 48.012  & 2.860   &   & 47.902  & 0.071   \\
HE 0043-0318     & 67.616  & 1.591     & 2002.12.02  & 01:30:55  & -26.459 & 68.302  & 0.967   &   & 70.191  & 2.082   &   & 68.637  & 1.020   \\
                 &         &           & 2003.01.16  & 01:44:09  & -29.463 & 66.376  & 1.124   &   & \nodata & \nodata &   & 66.376  & 1.124   \\
WD 0047-524      & 23.980  & 0.368     & 2002.07.27  & 06:21:40  & 3.620   & 24.533  & 0.795   &   & 22.385  & 1.627   &   & 24.119  & 1.198   \\
                 &         &           & 2002.09.14  & 03:12:24  & -10.877 & 24.175  & 0.716   &   & 19.757  & 1.671   &   & 23.488  & 2.263   \\
HS 0047+1903     & 27.280  & 0.926     & 2002.09.27  & 05:30:35  & 10.719  & 27.042  & 1.183   &   & 29.084  & 3.262   &   & 27.280  & 0.926   \\
WD 0048-544      & 22.199  & 0.822     & 2002.07.27  & 06:30:11  & 2.495   & 22.080  & 1.035   &   & 26.192  & 2.158   &   & 22.850  & 2.268   \\
                 &         &           & 2002.09.14  & 03:21:12  & -11.578 & 23.188  & 0.995   &   & 20.319  & 0.944   &   & 21.679  & 2.026   \\
WD 0048+202      & 35.980  & 0.587     & 2002.09.19  & 04:52:42  & 15.032  & 34.901  & 2.313   &   & 36.145  & 1.517   &   & 35.771  & 0.807   \\
                 &         &           & 2002.09.27  & 05:16:50  & 11.230  & 36.496  & 1.197   &   & 39.764  & 2.555   &   & 37.085  & 1.776   \\
                 &         &           & 2002.12.02  & 02:09:09  & -18.947 & 36.612  & 1.150   &   & 31.783  & 2.525   &   & 35.783  & 2.575   \\
HE 0049-0940     & 27.763  & 0.192     & 2002.09.26  & 08:39:23  & 0.509   & 27.896  & 1.001   &   & 26.624  & 2.178   &   & 27.675  & 0.682   \\
                 &         &           & 2002.09.27  & 06:27:15  & 0.251   & 28.308  & 0.783   &   & 26.327  & 1.736   &   & 27.973  & 1.049   \\
WD 0050-332      & 35.597  & 1.801     & 2002.07.27  & 06:44:31  & 14.143  & 35.763  & 4.027   &   & 36.931  & 5.721   &   & 36.150  & 0.777   \\
                 &         &           & 2002.09.15  & 03:26:56  & -3.174  & 31.345  & 2.874   &   & \nodata & \nodata &   & 31.345  & 2.874   \\
                 &         &           & 2002.09.25  & 06:49:51  & -7.716  & 32.107  & 2.952   &   & \nodata & \nodata &   & 32.107  & 2.952   \\
WD 0052-147      & 56.344  & 1.331     & 2002.09.26  & 08:54:09  & -1.164  & 58.289  & 1.884   &   & \nodata & \nodata &   & 58.289  & 1.884   \\
                 &         &           & 2002.09.27  & 06:41:44  & -1.422  & 56.297  & 2.405   &   & 54.873  & 3.785   &   & 55.888  & 0.912   
\enddata
\end{deluxetable*}
\clearpage
\end{landscape}

\clearpage

\LongTables 
\begin{landscape}
\begin{center}
\begin{deluxetable*}{cccccccccccccc}
\tablewidth{0pt}
\tabletypesize{\scriptsize}
\tablecaption{Apparent Velocity Measurements for Thick Disk DAs\label{table_thick}}
\tablehead{
\colhead{} & \multicolumn{2}{c}{Adopted} & \colhead{} & \colhead{} & \colhead{LSR} & \multicolumn{2}{c}{H$\alpha$} & \colhead{} & \multicolumn{2}{c}{H$\beta$} & \colhead{} & \multicolumn{2}{c}{Observation}
\\
\cline{2-3} \cline{7-8} \cline{10-11} \cline{13-14} \vspace{-0.12in}
\\
\colhead{} & \colhead{$v_{\rm app}$} & \colhead{$\delta v_{\rm app}$} & \colhead{Date} & \colhead{Time} & \colhead{Correction} & \colhead{$v_{\rm app}$} & \colhead{$\delta v_{\rm app}$} & \colhead{} & \colhead{$v_{\rm app}$} & \colhead{$\delta v_{\rm app}$} & \colhead{} & \colhead{$v_{\rm app}$} & \colhead{$\delta v_{\rm app}$} 
\\
\colhead{Target} & \colhead{(km s$^{-1}$)} & \colhead{(km s$^{-1}$)} & \colhead{(UT)} & \colhead{(UT)} & \colhead{(km s$^{-1}$)} & \colhead{(km s$^{-1}$)} & \colhead{(km s$^{-1}$)} & \colhead{} & \colhead{(km s$^{-1}$)} & \colhead{(km s$^{-1}$)} & \colhead{} & \colhead{(km s$^{-1}$)} & \colhead{(km s$^{-1}$)}
}
\startdata
WD 0158-227   & -12.279 & 2.795 & 2002.09.20 & 03:40:17 & -5.823  & -10.305 & 1.125 & & \nodata & \nodata & & -10.305 & 1.125  \\
              &         &       & 2002.09.27 & 07:46:45 & -9.161  & -14.259 & 1.127 & & \nodata & \nodata & & -14.259 & 1.127  \\
WD 0204-233   & 82.384  & 0.304 & 2000.07.15 & 07:45:18 & 8.895   & 82.192  & 0.944 & & 80.422  & 1.662   & & 81.760  & 1.074  \\
              &         &       & 2000.07.17 & 08:47:06 & 8.750   & 82.364  & 1.977 & & 83.186  & 5.477   & & 82.459  & 0.371  \\
WD 0255-705   & 47.623  & 0.527 & 2000.08.03 & 09:55:52 & -35.536 & 48.181  & 1.547 & & \nodata & \nodata & & 48.181  & 1.547  \\
              &         &       & 2000.08.05 & 08:51:01 & -35.778 & 47.374  & 1.034 & & \nodata & \nodata & & 7.374   & 1.034  \\
WD 0352+052   & -86.636 & 2.300 & 2002.03.01 & 01:09:46 & -35.737 & -87.408 & 1.401 & & \nodata & \nodata & & -87.408 & 1.401  \\
              &         &       & 2002.09.13 & 09:27:33 & 19.231  & -84.384 & 1.248 & & -79.498 & 2.216   & & -83.208 & 2.954  \\
HE 0409-5154  & 23.915  & 6.616 & 2000.09.15 & 09:10:28 & -37.306 & 22.243  & 2.047 & & 20.299  & 1.634   & & 21.056  & 1.340  \\
              &         &       & 2001.09.13 & 09:26:42 & -37.020 & 30.762  & 1.257 & & 34.543  & 2.406   & & 31.572  & 2.194  \\
HE 0416-1034  & 44.087  & 0.191 & 2000.12.17 & 06:13:24 & -38.231 & 43.785  & 1.509 & & 45.786  & 3.788   & & 44.059  & 0.973  \\
              &         &       & 2001.01.15 & 03:03:25 & -48.043 & 40.481  & 2.058 & & 47.436  & 1.631   & & 44.754  & 4.787  \\
HE 0452-3444  & -10.387 & 0.086 & 2000.12.13 & 06:27:39 & -44.215 & -9.733  & 1.464 & & -13.936 & 3.554   & & -10.343 & 2.094  \\
              &         &       & 2001.01.15 & 02:44:23 & -51.955 & -9.238  & 2.210 & & -13.956 & 3.712   & & -10.472 & 2.932  \\
HE 0508-2343  & 79.982  & 2.028 & 2001.04.07 & 00:23:02 & -59.490 & 79.124  & 1.786 & & 80.903  & 1.307   & & 80.283  & 1.198  \\
              &         &       & 2001.04.09 & 00:48:48 & -59.118 & 76.884  & 1.817 & & 68.789  & 1.966   & & 73.155  & 5.706  \\
WD 0732-427   & 36.295  & 0.558 & 2001.04.09 & 01:16:13 & -68.414 & 36.003  & 0.756 & & 37.866  & 1.433   & & 36.409  & 1.087  \\
              &         &       & 2001.05.03 & 23:55:10 & -69.286 & 33.730  & 1.190 & & 40.950  & 2.679   & & 34.920  & 3.788  \\
HS 0820+2503  & 41.390  & 4.704 & 2003.02.18 & 02:42:33 & -33.578 & 41.390  & 4.704 & & \nodata & \nodata & & 41.390  & 4.704  \\
HE 1124+0144  & 46.460  & 0.144 & 2000.07.01 & 23:28:45 & -55.699 & 45.879  & 1.394 & & 48.423  & 2.799   & & 46.384  & 1.435  \\
              &         &       & 2000.07.02 & 23:34:25 & -55.550 & 47.083  & 1.185 & & 42.748  & 3.337   & & 46.598  & 1.932  \\
WD 1152-287   & 51.828  & 1.745 & 2000.07.11 & 00:57:00 & -70.466 & 50.760  & 1.594 & & \nodata & \nodata & & 50.760  & 1.594  \\
              &         &       & 2000.07.14 & 23:01:27 & -70.053 & 53.254  & 1.841 & & \nodata & \nodata & & 53.254  & 1.841  \\
WD 1323-514   & -40.353 & 0.770 & 2001.05.15 & 01:29:38 & -42.147 & -40.093 & 0.793 & & -41.065 & 2.400   & & -40.189 & 0.409  \\
              &         &       & 2001.06.08 & 02:12:18 & -50.778 & -42.707 & 1.018 & & -40.445 & 1.821   & & -42.168 & 1.362  \\
WD 1334-678   & 22.946  & 0.649 & 2000.07.30 & 00:47:19 & -58.211 & 22.384  & 1.473 & & 24.018  & 3.072   & & 22.689  & 0.900  \\
              &         &       & 2001.05.15 & 01:42:37 & -40.116 & 24.128  & 1.309 & & 20.157  & 4.143   & & 23.767  & 1.612  \\
HS 1338+0807  & 64.191  & 2.405 & 2001.08.18 & 23:52:07 & -28.395 & 64.191  & 2.405 & & \nodata & \nodata & & 64.191  & 2.405  \\
WD 1410+168   & 15.184  & 2.748 & 2002.04.23 & 06:08:33 & 7.586   & 15.753  & 0.911 & & 8.557   & 3.107   & & 15.184  & 2.748  \\
WD 1426-276   & 55.800  & 0.559 & 2000.07.05 & 03:56:09 & -43.083 & 55.742  & 1.182 & & 55.448  & 3.136   & & 55.705  & 0.137  \\
              &         &       & 2000.07.06 & 02:38:26 & -43.227 & 57.693  & 1.045 & & 56.772  & 1.755   & & 57.452  & 0.572  \\
HS 1432+1441  & 84.966  & 3.366 & 2001.08.16 & 00:01:42 & -14.425 & 82.686  & 1.642 & & 77.964  & 3.260   & & 81.730  & 2.683  \\
              &         &       & 2001.08.21 & 00:15:53 & -13.555 & 85.505  & 1.995 & & 88.323  & 2.297   & & 86.717  & 1.973  \\
WD 1507+021   & 50.907  & 3.076 & 2002.06.18 & 01:25:19 & -7.271  & 52.338  & 2.198 & & 47.601  & 3.340   & & 50.907  & 3.076  \\
WD 1614-128   & 95.682  & 0.052 & 2000.06.06 & 06:16:38 & 5.840   & 94.870  & 1.239 & & 99.368  & 2.581   & & 95.713  & 2.482  \\
              &         &       & 2000.06.08 & 02:20:56 & 5.346   & 94.931  & 1.088 & & 102.053 & 3.282   & & 95.637  & 3.009  \\
WD 1716+020   & 10.643  & 1.291 & 2002.04.23 & 09:03:16 & 48.550  & 11.038  & 0.846 & & 8.531   & 1.956   & & 10.643  & 1.291  \\
WD 1834-781   & 60.781  & 0.111 & 2000.07.06 & 04:36:43 & -35.096 & 60.857  & 0.919 & & 60.355  & 2.395   & & 60.792  & 0.237  \\
              &         &       & 2000.07.13 & 05:04:33 & -37.059 & 59.611  & 1.028 & & 62.217  & 1.793   & & 60.255  & 1.589  \\
WD 1952-206   & 62.701  & 0.113 & 2000.07.06 & 05:09:15 & 29.248  & 62.167  & 0.870 & & 63.309  & 2.141   & & 62.329  & 0.563  \\
              &         &       & 2000.07.13 & 04:43:35 & 25.831  & 62.683  & 0.956 & & 62.931  & 2.356   & & 62.718  & 0.122  \\
WD 2029+183   & -97.198 & 1.373 & 2002.04.24 & 09:34:58 & 74.163  & -98.949 & 2.029 & & -96.770 & 1.646   & & -97.635 & 1.507  \\
              &         &       & 2002.08.05 & 04:50:45 & 51.952  & -93.455 & 1.652 & & -98.577 & 2.467   & & -95.041 & 3.349  \\
WD 2322-181   & 39.763  & 1.328 & 2000.07.13 & 06:16:26 & 38.666  & 42.172  & 1.462 & & 38.142  & 2.548   & & 41.173  & 2.459  \\
              &         &       & 2000.07.16 & 05:46:18 & 37.813  & 39.731  & 1.648 & & 36.877  & 3.217   & & 39.138  & 1.638  \\
WD 2350-083   & 84.801  & 0.992 & 2002.07.11 & 09:56:33 & 49.133  & 84.652  & 1.390 & & 85.845  & 2.341   & & 84.963  & 0.740  \\
              &         &       & 2002.09.13 & 07:51:37 & 24.339  & 80.737  & 1.489 & & 86.827  & 3.323   & & 81.756  & 3.214  
\enddata
\end{deluxetable*}
\end{center}
\clearpage
\end{landscape}

\begin{deluxetable*}{cccccccc}
\tablewidth{0pt}
\tabletypesize{\footnotesize}
\tablecaption{Mean Apparent Velocities\label{results}}
\tablehead{
\colhead{} & \colhead{} & \colhead{$\langle v_{\rm app}\rangle$} & \colhead{$\delta\langle v_{\rm app}\rangle$} & \colhead{$\sigma _{v_{\rm app}}$} & \colhead{$\langle \delta$$v_{\rm app}\rangle$} & \colhead{$\langle M/R\rangle$} & \colhead{$\delta\langle M/R\rangle$}
\\
\colhead{Sample} & \colhead{\# of WDs} & \colhead{(km s$^{-1}$)} & \colhead{(km s$^{-1}$)} & \colhead{(km s$^{-1}$)} & \colhead{(km s$^{-1}$)} &\colhead{(M$_\odot$/R$_\odot$)} & 
\colhead{(M$_\odot$/R$_\odot$)}
}
\startdata
Main & 449 & 32.57 & 1.17 & 24.84 & 1.51 & 51.19 & 1.84
\\
H$\alpha$ & 449 & 32.69 & 1.18 & 24.87 & 1.78 & 51.37 & 1.85
\\
H$\beta$ & 372 & 31.47 & 1.32 & 25.52 & 3.17 & 49.45 & 2.07
\\
Thick & 26 & 32.90 & 9.59 & 48.99 & 1.57 & 51.70 & 15.07
\enddata
\end{deluxetable*}

\begin{deluxetable*}{cccccccc}
\tablewidth{0pt}
\tablecaption{Mean Masses\label{table_masses}}
\tablehead{
\colhead{} & \colhead{} & \colhead{$\langle v_{\rm app}\rangle$} & \colhead{$\delta$\,$\langle v_{\rm app}\rangle$} & \colhead{$\langle T_{\rm eff}\rangle$} & \colhead{$\sigma_{T_{\rm eff}}$} & \colhead{$\langle M\rangle$} & \colhead{$\delta$\,$\langle M\rangle$}
\\
\colhead{Sample} & \colhead{\# of WDs} & \colhead{(km s$^{-1}$)} & \colhead{(km s$^{-1}$)} & \colhead{(K)} & \colhead{(K)} & \colhead{(M$_\odot$)} & 
\colhead{(M$_\odot$)}
}
\startdata
Main & 449 & 32.57 & 1.17 & 19400 & 9950 & 0.647 & $^{+0.013}_{-0.014}$
\\
Thick & 26 & 32.90 & 9.59 & 19960 & 11060 & 0.652 & $^{+0.097}_{-0.119}$
\\
Hot\tablenotemark{a} & 366 & 31.61 & 1.22 & 21670 & 9700 & 0.640 & 0.014
\\
Cool\tablenotemark{a} & 75 & 37.50 & 3.59 & 9950 & 1090 & 0.686 & $^{+0.035}_{-0.039}$
\enddata
\tablenotetext{a}{``Hot'' refers to WDs with $T_{\rm eff}>12000$\,K and ``cool'' to WDs with $12000$\,K\,$>T_{\rm eff}>7000$\,K.}
\end{deluxetable*}

\centering{
\begin{deluxetable*}{lrlllccc}
\tablewidth{0pt}
\tabletypesize{\scriptsize}
\tablecaption{Mean DA Masses From Selected Previous Studies\label{previous}}
\tablehead{
\colhead{} & \colhead{} & \colhead{$\langle M\rangle$} & \colhead{$\delta$\,$\langle M\rangle$} & \colhead{$\sigma_M$} & \colhead{} & \colhead{Assumed} & \colhead{}
\\
\colhead{Study} & \colhead{\# of WDs} & \colhead{(M$_\odot$)} & \colhead{(M$_\odot$)} & \colhead{(M$_\odot$)} & \colhead{Method} & \colhead{H-Layer\tablenotemark{a}} & \colhead{Notes}
}
\startdata
\citet{koester79} & 122 & 0.58 & 0.10 & 0.12\tablenotemark{b} & Photo & Thin/No & 
\\
\citet{koester87} & 9 & 0.58 & \nodata & 0.11 & GRS & Thin/No & CPM WDs
\\
\citet{mcmahan89} & 50 & 0.523 & 0.014 & \nodata & Spectro & Thin/No &
\\
\citet{wegner91} & 35 & 0.63 & 0.03 & \nodata & GRS & Thin/No & CPM WDs
\\
\citet{bergeron92b} & 129 & 0.562 & \nodata & 0.137 & Spectro & Thin/No & $T_{\rm eff}\gtrsim14000$\,K
\\
\citet{bragaglia95} & 42 & 0.609 & \nodata & 0.157 & Spectro & Thin/No & $T_{\rm eff}>12000$\,K
\\
\citet{bergeron95} & 129 & 0.590 & \nodata & 0.134 & Spectro & Thick & Revised \citet{bergeron92b}
\\
 & & & & & & & w/ thick H-layers
\\
\citet{reid96} & 34 & 0.583 & 0.078 & \nodata & GRS & Thick & CPM WDs
\\
\citet{vennes97} & 110 & 0.56* & \nodata & \nodata & Spectro & Thin/No & $75000$\,K $\gtrsim T_{\rm eff}\gtrsim25000$\,K
\\
\citet{finley97} & 174 & 0.570* & \nodata & 0.060* & Spectro & Thick & $T_{\rm eff}\gtrsim25000$\,K
\\
 & & & & & & & some w/ cool companions
\\
\citet{silvestri01} & 41 & 0.68 & 0.04 & \nodata & GRS & Thick & CPM WDs
\\
\citet{madej04} & 1175 & 0.562* & \nodata & \nodata & Spectro & Thick & $T_{\rm eff}\ge 12000$\,K
\\
\citet{liebert05} & 298 & 0.603 & \nodata & 0.134 & Spectro & Thick & $T_{\rm eff}>13000$\,K
\\
 & & 0.572* & \nodata & 0.188 & & &
\\
\citet{kepler07} & 1859 & 0.593 & 0.016 & \nodata & Spectro & Thick & $T_{\rm eff}>12000$\,K
\\
\citet{tremblay09} & $\sim250$ & 0.649 & \nodata & \nodata & Spectro & Thick & $40000$\,K $>T_{\rm eff}>12000$\,K
\\
 & & & & & & & overlap w/ Liebert et al.
\\
\citet{koester09b}\tablenotemark{c} & 606\tablenotemark{d} & 0.567\tablenotemark{e} & 0.002\tablenotemark{e} & 0.142\tablenotemark{e} & Spectro & Thick & SPY
\\
\citet{koester09b}\tablenotemark{c} & 512\tablenotemark{d} & 0.580\tablenotemark{e} & 0.002\tablenotemark{e} & 0.136\tablenotemark{e} &  &  & SPY non-binary thin disk WDs
\\
\citet{koester09b} overlap\tablenotemark{c} & 441 & 0.575\tablenotemark{e} & 0.002\tablenotemark{e} & 0.128\tablenotemark{e} & & & 
\\
This Work & 449 & 0.647 & $^{+0.013}_{-0.014}$ & \nodata & GRS & Thick & SPY non-binary thin disk WDs
\enddata
\tablecomments{Masses marked with an asterisk are peaks/widths of mass distributions from Gaussian fitting.}
\tablenotetext{a}{Hydrogen layer mass used in mass-radius relation from evolutionary models. ``Thick'' corresponds to $M_{\rm H}/M_\star\approx 10^{-4}$ and ``Thin/No'' to $M_{\rm H}/M_\star\lesssim10^{-8}$ or no hydrogen layer.}
\tablenotetext{b}{Two-thirds of the stars are within 0.12\,M$_\odot$.}
\tablenotetext{c}{Masses do not appear in this reference.  We compute masses from the published values of log\,$g$ and $T_{\rm eff}$ using the mass-radius relation from evolutionary models.}
\tablenotetext{d}{Excludes double degenerates.}
\tablenotetext{e}{We compute these means, uncertainties, and standard deviations (see Section \ref{masses}).}
\end{deluxetable*}
}

\end{document}